\documentclass[useAMS,usenatbib,twocolumn]{mnras}
\usepackage{epsfig,amssymb}
\usepackage{graphicx,amsmath,multicol}
\usepackage{multirow}
\usepackage{caption}
\expandafter\def\csname ver@subfig.sty\endcsname{}
\usepackage{placeins}
\usepackage{subfig}
\usepackage{mathrsfs}
\usepackage{physics}

\usepackage[utf8]{inputenc}
\usepackage{setspace}
\usepackage{mathtools}
\usepackage{grffile}
\usepackage{amsmath}
\usepackage{longtable}
\usepackage{graphics}
\usepackage{verbatim}
\usepackage[usenames]{color}
\usepackage{float}

\newcommand{\kms}{km\,s$^{-1}$} 
\newcommand{\Ts}   {\ensuremath{{\it T}_{\rm s}}}

\newcommand{\Tk}   {\ensuremath{{\it T}_{\rm k}}}
\newcommand{\Tb}   {\ensuremath{{\it T}_{\rm B}}}
\newcommand{\Tbtot}   {\ensuremath{{\it T}_{\rm B,tot}}}

\newcommand{\nhi}  {\ensuremath{N}_{\rm HI}}
\newcommand{\nhiot}{\ensuremath{N_{\rm HI, OT}}}
\newcommand{\nhits}{\ensuremath{N_{\rm HI, ISO}}}
\newcommand{\nhies}{\ensuremath{N_{\rm HI, E}}}

\title[H\,{\sc i} column density from emission spectra]{On estimating the atomic hydrogen column density from the H\,{\sc i} $21$ cm emission spectra}
\author[Saha, Roy \& Bhattacharya]{Preetha Saha,$^{1}$\thanks{E-mail: preethasaha06@gmail.com} Nirupam Roy$^{2}$ and Mukul Bhattacharya$^{3}$\\~\\
$^{1}$ Department of Physics and Centre for Theoretical Studies, Indian Institute of Technology, Kharagpur 721302, India\\
$^{2}$ Department of Physics, Indian Institute of Science, Bangalore 560012, India\\
$^{3}$ Department of Physics, University of Texas at Austin, Austin, TX 78712, USA}

\begin{document}
\date{Accepted . Received ; in original form}
\pagerange{\pageref{firstpage}--\pageref{lastpage}} 
\pubyear{2018}

\maketitle

\label{firstpage}

\begin{abstract}
The $21$ cm hyperfine transition of the atomic hydrogen (H\,{\sc i}) in ground 
state is a powerful probe of the neutral gas content of the universe. This 
radio frequency transition has been used routinely for decades to observe, both 
in emission and absorption, H\,{\sc i} in the Galactic interstellar medium as 
well as in extragalactic sources. In general, however, it is not trivial to 
derive the physically relevant parameters like temperature, density or column 
density from these observations. Here, we have considered the issue of column 
density estimation from the H\,{\sc i} $21$ cm emission spectrum for sightlines 
with a non-negligible optical depth and a mix of gas at different temperatures. 
The complicated radiative transfer and a lack of knowledge about the relative 
position of gas clouds along the sightline often make it impossible to uniquely 
separate the components, and hinders reliable estimation of column densities in 
such cases. Based on the observed correlation between the $21$ cm brightness 
temperature and optical depth, we propose a method to get an unbiased estimate 
of the H\,{\sc i} column density using only the $21$ cm emission spectrum. This 
formalism is further used for a large sample to study the spin temperature of 
the neutral interstellar medium.
\end{abstract}


\section{Introduction}
\label{sec:intro}

Atomic hydrogen (H\,{\sc i}) is the main constituent of the diffuse neutral 
interstellar medium (ISM). The H\,{\sc i} $21$ cm radio frequency transition 
between the two hyperfine levels of the ground state (at $1420.4057517$ MHz) 
is used extensively to study the ISM of the Milky Way, the ISM of other nearby 
galaxies as well as redshifted cosmological signal from neutral gas in the 
distant universe \citep[e.g.][]{clark62,field65,field69,crovisierdickey83,
walter08}.

The $21$ cm spectral line may be observed either in emission or in absorption 
(against suitable background continuum sources). The populations of the two 
hyperfine levels are related by the spin temperature $\Ts$, and decide the 
relative strength of emission and absorption. The emission spectrum gives us 
the specific intensity $I_\nu$. In the Rayleigh-Jeans regime (i.e. $h\nu<<kT$), 
this is conveniently expressed as brightness temperature $\Tb = I_{\nu}c{^2} / 
2k\nu^2$ where $k$ is Boltzmann's constant, $\nu$ is frequency and $c$ is the 
speed of light. The absorption spectrum, on the other hand, provides the 
H\,{\sc i} $21$ cm optical depth $\tau$ that depends on the linear absorption 
coefficient $\kappa_\nu$ which, in turn, depends on $\Ts$ and the density of 
the H\,{\sc i}.

The direct observables in H\,{\sc i} $21$ cm absorption and emission studies 
are the Doppler shift velocity of the spectral line $V_c$, width of the line 
due to thermal and non-thermal broadening $\Delta V$, $\Tb(V)$ and $\tau(V)$ 
(from emission and absorption studies respectively) over the velocity range of 
the line profile. While the central velocity $V_c$ is useful in studying the 
dynamics of the ISM; the other quantities, in combination, can be used to 
estimate physical properties like the temperature, the density or the column 
density of the gas in certain conditions and under certain assumptions.

In this paper, we carefully reconsider the issue of column density measurements 
using H\,{\sc i} $21$ cm studies. In the general case, when the sightline under 
consideration passes either through a mix of different phases of gas or, 
equivalently, through multiple ``clouds'' at different temperatures, it is not 
straightforward to infer the column density from the observed absorption or 
emission spectrum. Moreover, for lines of sight with higher value of $\tau$, 
the emission spectrum can be used to get the optically thin limit of the column 
density. This measurement is significantly biased as the optically thin limit 
underestimates the column density. Alternatively, one may use both emission and 
absorption spectra to get an unbiased estimate of the column density. However, 
absorption studies need suitable background continuum sources for the same or a 
nearby sightline, and may not always be feasible to carry out. We suggest here 
to utilize a physically motivated, as well as observationally established 
correlation between $\Tb$ and $\tau$, to derive an unbiased H\,{\sc i} column 
density from only the observed emission spectrum. In this paper, we describe 
the formalism in Section \ref{sec:nhcol}, and outline the method in Section 
\ref{sec:ttauc}. In Section \ref{sec:reslt} we show the application of this 
method. Some possible limitations of this method are discussed in Section 
\ref{sec:dscus} along with conclusions.

\section{H\,{\sc i} column density measurement}
\label{sec:nhcol}

Considering an isothermal cloud, the atomic hydrogen column density $\nhi$ may 
be written as
\begin{equation}
\label{eqn:nhitstau}
\nhi = (1.823 \times 10^{18} \: {\rm cm}^{-2}) \int \Ts \: \tau \: {dV},
\end{equation}
where $\Ts$ is in K, velocity interval ${dV}$ is in \kms, and the integral is 
over the velocity range of the cloud \citep{kulkarniheiles87,dickey90}. Please 
note that velocity dependence of $\Ts$ and $\tau$ are not shown explicitly. One 
can measure $\tau$ from absorption studies towards suitable continuum sources. 
$\Ts$ can also be derived by combining $\Tb$ and $\tau$ using the relation
\begin{equation}
\label{eqn:tbts}
\Tb = \Ts \left[ 1 - \exp(-\tau) \right],
\end{equation}
where $\Tb$ is measured from the H\,{\sc i} $21$ cm emission studies. Thus, 
from equations~(\ref{eqn:nhitstau}) and (\ref{eqn:tbts}), $\nhi$ for a cloud 
under the isothermal assumption \citep{dickey82} is
\begin{equation}
\label{eqn:nhitbtau}
\nhi = (1.823 \times 10^{18} \: {\rm cm}^{-2}) \int \frac{\tau\: \Tb} {\left[ 1 - \exp(-\tau) \right]}\: {dV}  
\end{equation}
in terms of direct observables $\Tb$ and $\tau$. For the optically thin limit 
($\tau << 1$), one may further simplify this to
\begin{equation}
\label{eqn:nhiot}
\nhi = (1.823 \times 10^{18} \: {\rm cm}^{-2}) \int \Tb\: {dV} 
\end{equation}
to estimate $\nhi$ only from the emission studies.

In reality, however, a given sightline will pass through a number of clouds (or 
a mix of gases) at different temperatures, and the optical depth, most often, 
is also not negligible. Even for $\tau \approx 0.2$ ($0.5$), $\nhi$ differs by 
$10\%$ ($30\%$) from the optically thin approximation. Thus, both equations 
(\ref{eqn:nhitbtau}) and (\ref{eqn:nhiot}) will not be readily applicable to 
estimate $\nhi$. Then, one can only measure $\Tbtot$ and $\tau_{\rm tot}$, 
i.e.~the combined total contribution of $\Tb$ and $\tau$ at a given velocity 
``channel'' by all the clouds along the sightline. Further, the complicated 
radiative transfer makes it impossible to uniquely separate the contributions 
to $\Tb$ from different components. In this case, we can either derive a lower 
limit of $\nhi$ using the optically thin approximation
\begin{equation}
\label{eqn:nhiot1}
\nhiot = (1.823 \times 10^{18} \: {\rm cm}^{-2}) \int \Tbtot\: {dV},
\end{equation}
or use the isothermal approximation to derive 
\begin{equation}
\label{eqn:nhits}
\nhits = (1.823 \times 10^{18} \: {\rm cm}^{-2}) \int \frac{\tau_{\rm tot} \: \Tbtot} {\left[ 1 - \exp(-\tau_{\rm tot}) \right]}\: {dV}.
\end{equation}

Extensive numerical simulations by \citet{chengalur13} have shown that 
$\nhiot$ grossly underestimates the true column density, whereas $\nhits$ is an 
unbiased estimator independent of gas temperature distribution or positions of 
clouds along the sightline. These results hold for $\nhi$ as high as $\le 
5\times 10^{23}$\, cm$^{-2}$ per $1$~km~s$^{-1}$ channel and $\tau$ $\le 1000$. 
Unfortunately, this still requires independent estimation of both $\Tb$ and 
$\tau$ from emission and absorption studies respectively. As it may not always 
be possible to find a suitable background continuum source to get the $21$ cm 
absorption spectrum, emission studies often can only provide the optically thin 
limit of $\nhi$. Any other indirect estimation of $\nhi$ from emission study is 
only possible under more assumptions, e.g. extrapolating optical depth from 
nearby lines of sight, that may often be unreliable (e.g. \citealt{heiles03b}, 
reported variation of $\tau$ by a factor as high as $2.5$ for few arcmin 
separation).

\section{Method using $\Tb-\tau$ correlation}
\label{sec:ttauc}

Here, we present a method for estimating $\nhi$ from only the 21 cm emission 
spectrum using an empirical $\Tb-\tau$ correlation. The $21$ cm optical depth 
is proportional to the H\,{\sc i} volume density $\rho$ and $\Ts^{-1}$ 
\citep{kulkarniheiles87}
\begin{equation}
\tau \propto \rho/\Ts\: ,
\end{equation}
where $\Ts$ is related to the population of the two hyperfine levels and is 
considered to be a good proxy of the kinetic temperature $\Tk$ for the cold gas 
\citep{liszt01}. At high enough densities in the cold phase, $\Ts$ is tightly 
coupled to $\Tk$ via collisions. At lower densities, collisions are less, and 
$\Ts$ is in general lower than $\Tk$. Here we assume a simple parametric 
relation between $\Tk$ and $\Ts$ of the form $\Ts \propto (\Tk)^\alpha$, where 
$\alpha \leq 1$. We also assume an equation of state relating $\rho$ and $\Tk$ 
of the form
\begin{equation}
\rho \propto (\Tk)^n
\end{equation}
where $n = 1/(1-\gamma)$ is the polytropic index and $\gamma$ is the adiabatic 
index. One can also consider $\alpha$ to be a function of $n$, but for 
simplicity, $\alpha$ is kept constant in this analysis. If the different phases 
of the ISM along a sightline are in rough thermal pressure equilibrium 
\citep{field65,field69}, then $n = -1$ (so that pressure $P \propto \rho \Tk$ 
is constant). In this case, the optical depth 
\begin{equation}
\label{eqn:eqn9}
\tau \propto \Ts^{-(1+\alpha)/\alpha} .
\end{equation}
Combining equation~(\ref{eqn:tbts}) and (\ref{eqn:eqn9}), we can write
\begin{equation}
\Tb \propto \tau^{-\alpha/(1+\alpha)}\:\left[1-\exp(-\tau)\right] .
\end{equation}

\begin{figure}
\begin{center}
\includegraphics[scale=0.33,angle=270.0]{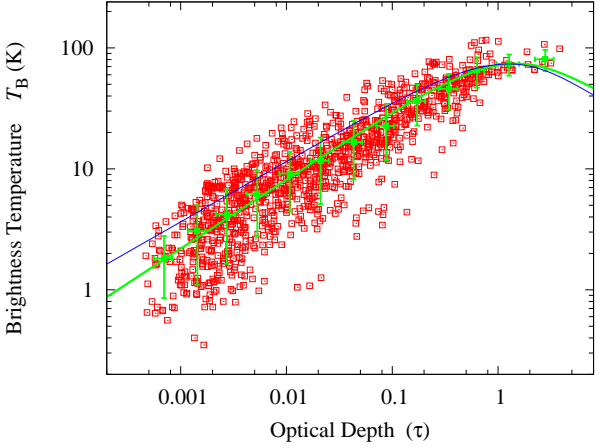}
\caption{Observed $\Tb-\tau$ correlation based on data from \citet{roy13a}. The 
open squares are from individual velocity channels and the filled squares with 
error bars are binned data. The thick and the thin lines are for $\alpha = 3/4$ 
and $1$ respectively.}
\label{fig:fig1} 
\end{center}
\end{figure}

Next, we validate this physically motivated, simple model using observational 
data. For this we have taken $\tau$ from the high spectral resolution and high 
sensitivity H\,{\sc i} absorption survey by \citet{roy13a}. This is an {\it 
ongoing} survey of the Galactic H\,{\sc i} 21 cm absorption using the Giant 
Metrewave Radio Telescope (GMRT) and the Westerbork Synthesis Radio Telescope 
(WSRT), with an optical depth RMS sensitivity of $\lesssim 10^{-3}$ per $1$ 
km~s$^{-1}$ channel. \citet{roy13a} have reported the initial results based on 
data for 32 lines of sight. The corresponding $\Tb$ values are taken from the 
LAB survey \citep{kalberla05}, and the observed $\tau$ is smoothed to a 
matching resolution of $\sim 1$ \kms. The $\Tb - \tau$ data covering more than 
three orders of magnitude in $\tau$ is shown in Figure \ref{fig:fig1}. The open 
square symbols are showing all $\Tb$ and $\tau$ from the individual velocity 
channels measured with $>3\sigma$ significance for both. The filled squares 
with the error bars are the binned data with $1\sigma$ uncertainty. Here we 
have shown the mean values, but the mean and the median values are very close 
to each other in all the bins. The thin line is the model for $\alpha = 1$ 
(i.e.~ $\Ts = \Tk$), while the thick line is for $\alpha = 3/4$. Both the 
models are normalized at the same value of $\tau = 1.27$ (the second highest 
bin in $\tau$). The data clearly show a fairly good agreement with the model 
where $\Ts \propto (\Tk)^{3/4}$, hence indicating the expected deviation of 
$\Ts$ from $\Tk$ at lower optical depths. Also note that the turn around $\tau 
= 1$ indicates a plausible peak $\Tb$ due to self-absorption. 

\begin{figure}
\begin{center}
\includegraphics[scale=0.33,angle=270.0]{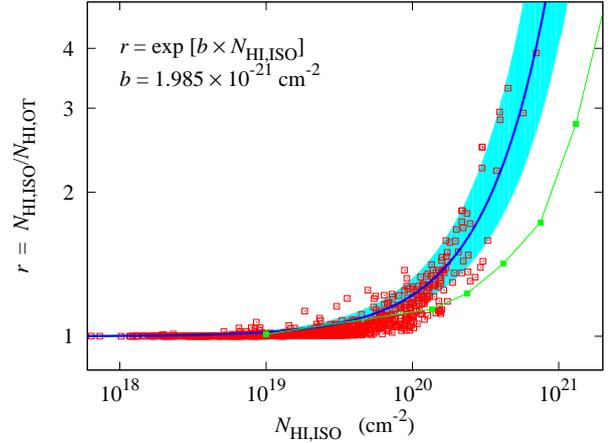}
\caption{The ratio of ``true'' to apparent $\nhi$ as a function of the true 
$\nhi$. As $\nhits$ is an unbiased estimate, we use that as a proxy for the 
true $\nhi$. The observed data from individual velocity channels are shown as 
open squares. The thick line and the shaded region are the best fit exponential 
function with a factor of two uncertainty in the exponent. The filled squares 
joined by thin line is taken from simulations by \citet{chengalur13}.}
\label{fig:fig2-3} 
\end{center}
\end{figure}

Based on this $\Tb - \tau$ correlation, we can define an estimator of $\nhi$ 
using only $\Tbtot$ for a velocity channel as
\begin{equation}
\label{eqn:nhies}
\nhies = (1.823 \times 10^{18} \: {\rm cm}^{-2})\: r(\Tbtot) \: \Tbtot dV \: , 
\end{equation}
where $dV$ is in ${\rm km~s}^{-1}$ and $r$ is a function of $\Tbtot$ or $\tau_{\rm tot}$
\begin{equation}
\label{eqn:eqnr}
r = \nhits/\nhiot = \frac{\tau_{\rm tot}}{\left[1-\exp(-\tau_{\rm tot})\right]}  \: .
\end{equation}

Figure \ref{fig:fig2-3} shows the observed ratio $r$ as a function of $\nhits$ 
per $\sim 1\:{\rm km\: s^{-1}}$ velocity channel. We have used a fiducial 
functional form 
\begin{equation}
\label{eqn:eqnr1}
r = r(\nhits) = 1.00\:\exp(b\:\nhits) \:.
\end{equation}
The best fit function and its variation for a factor of two change in the 
exponent $b$ are also shown in Figure \ref{fig:fig2-3}. This functional form 
can now be used to iteratively solve equations~(\ref{eqn:nhies}) and 
(\ref{eqn:eqnr}) to get $\nhies$ for the unit width velocity channel. To get 
the total $\nhi$, $\nhies$ should be summed over the full velocity range of the 
emission spectra. We have implemented this in a standard C code to estimate 
$\nhi$ from emission line, and the results are shown in the next section.

\section{Results and applications}
\label{sec:reslt}

\begin{figure}
\begin{center}
\includegraphics[scale=0.33,angle=270.0]{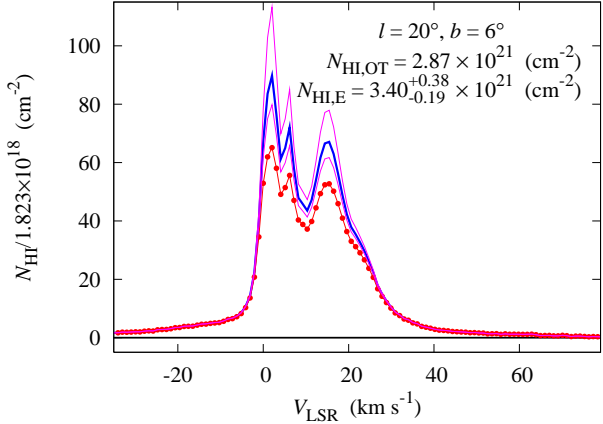}
\caption{Example H\,{\sc i} $21$ cm emission spectra from the LAB survey 
showing $\nhi$ per $1$ \kms channel. The filled circles joined by line (the 
thick line) are $\nhi$ before (after) correction for absorption (i.e.~$\nhiot$ 
and $\nhies$ respectively). The uncertainty in $\nhies$ (shown by the pair of 
thin lines) are for changing the exponent $b$ in equation~(\ref{eqn:eqnr1}) by 
a factor of two.}
\label{fig:fig4-5} 
\end{center}
\end{figure}

This formalism to estimate $\nhi$ from the $21$ cm emission spectra is applied 
to archival data from the LAB survey. In Figure \ref{fig:fig4-5}, an example 
spectrum is shown to demonstrate the change in the estimated column density 
$\nhies$ from the optically thin column density $\nhiot$. The observed $\Tb = 
\nhiot/1.823\times10^{18} \: {\rm cm}^{-2}$ per $1\:{\rm km\: s^{-1}}$ velocity 
channel is shown as filled points joined by a line. The corrected estimate of 
$\nhies$ per $1\:{\rm km\: s^{-1}}$ channel is shown as a thick line, and a 
pair of thin lines denote a factor of two uncertainty of the exponent $b$ in 
equation (\ref{eqn:eqnr1}). For the example sightline ($l = 20^\circ$, $b = 
6^\circ$), the $\nhiot$ value is $2.87 \times 10^{21} \: {\rm cm}^{-2}$, 
whereas the corrected $\nhi$ value is $\nhies = 3.40\times 10^{21} \: {\rm 
cm}^{-2}$. The error in $\nhi$ due to uncertainty in $\Tb$ is very small ($\sim 
0.07\: {\rm K} = 1.3\times 10^{17} \: {\rm cm}^{-2}$ per unit velocity 
interval). For an uncertainty in $b$ as large as a factor of two, the estimated 
$\nhies$ changes by $\lesssim 20\%$ only.

\begin{figure}
\begin{center}
\includegraphics[scale=0.33,angle=270.0]{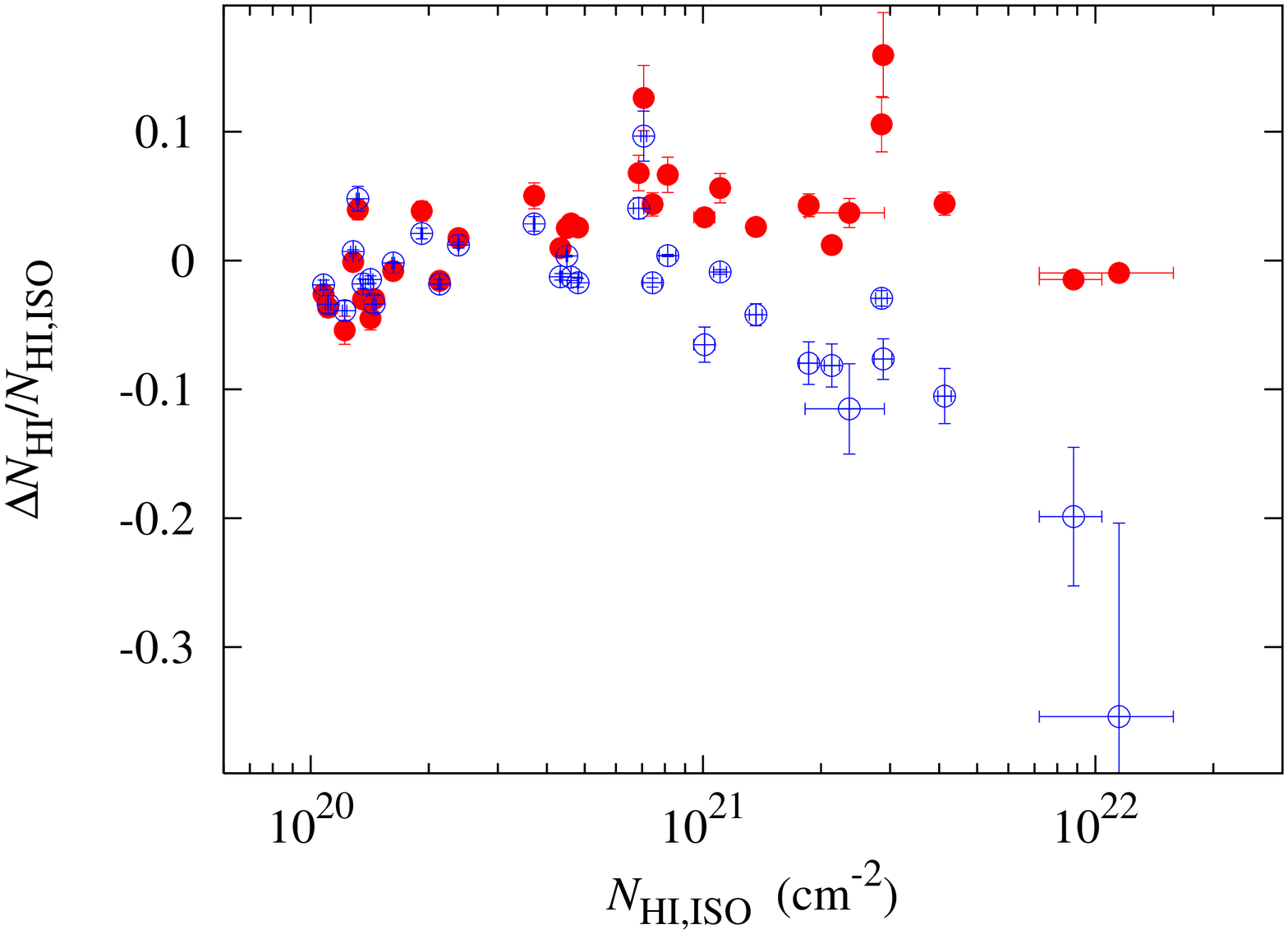}
\caption{A comparison of different $\nhi$ for \citet{roy13a} sample. The filled 
circles show the fractional difference between $\nhies$ and $\nhits$, and the 
open circles show the fractional difference between $\nhies$ and $\nhiot$. The 
error bars include an assumed $20\%$ uncertainty in $\nhies$.}
\label{fig:fig6} 
\end{center}
\end{figure}

Next, we use the sample of \citet{roy13a} to compare $\nhies$ and $\nhits$, and 
to check if $\nhies$ is indeed an unbiased estimator as well. Please note that 
the $\Tb-\tau$ correlation used for this formalism is also from the same 
sample. However, $\tau$ varies for the sample by more than three orders of 
magnitude, and the observed correlation is between the averaged quantities. So, 
there is no {\it a priori} reason to expect the two column densities to match 
closely for the individual lines of sight. Figure \ref{fig:fig6} shows the 
fractional deviation of $\nhies$ from $\nhits$, $(\nhies-\nhits)/\nhits$, for 
this sample (filled circles with error bars). A similar fractional deviation 
between $\nhiot$ and $\nhits$ is also shown (open circles with error bars) for 
comparison. At lower $\nhi$, all the estimates agree with each other. However, 
at higher $\nhi$, $\nhies$ matches better with $\nhits$. This ascertains that 
$\nhies$ is an useful and unbiased estimator of $\nhi$ even when no absorption 
measurement is available.

We further extend this analysis to a larger sample for which both emission and 
absorption measurements are reported in the literature. However, in many cases, 
the velocity resolution of the data is coarse, and thus $\nhits$ can not be 
computed reliably. One can, however, still estimate $\nhies$, and combine it 
with the integrated optical depth from the literature to get the average $\Ts$ 
for the sightline. This is effectively the column density weighted harmonic 
mean of $\Ts$ ($\langle\Ts\rangle$) of different components along the sightline 
\citep{kulkarniheiles87}. The estimator is then applied to a sample of 318 
sightlines, compiled from various H~{\sc i} absorption surveys after excluding 
non-detections and common sources: \citet[][87 sources, spectral resolution 
$\Delta V = 1.55$ km~s$^{-1}$]{dickey83}, \citet[][78 sources, $\Delta V = 
0.16$ km~s$^{-1}$]{heiles03b}, \citet[][102 sources, $\Delta V = 3.3$ 
km~s$^{-1}$]{mohan04}, \citet[][104 sources, $\Delta V = 0.1$ 
km~s$^{-1}$]{liszt10}, and \citet[][see above for details]{roy13a}. These 
sightlines have the observed and interpolated $\nhi$ in the range $\sim 8 
\times10^{19}$ ${\rm cm}^{-2}$ to $2\times10^{22}$ ${\rm cm}^{-2}$. 

\begin{figure*}
\begin{center}
\includegraphics[scale=0.33,angle=270.0]{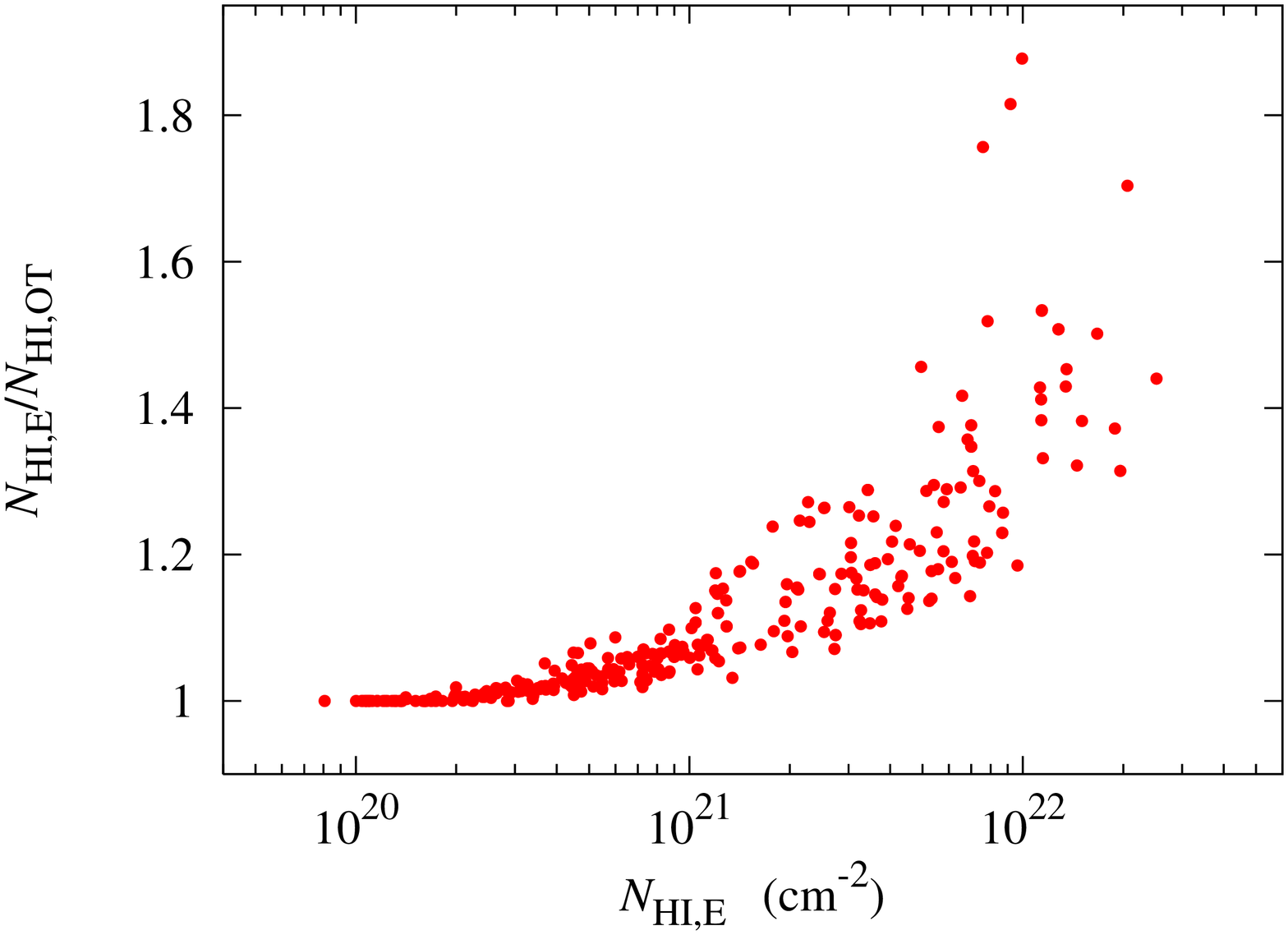}\includegraphics[scale=0.33,angle=270.0]{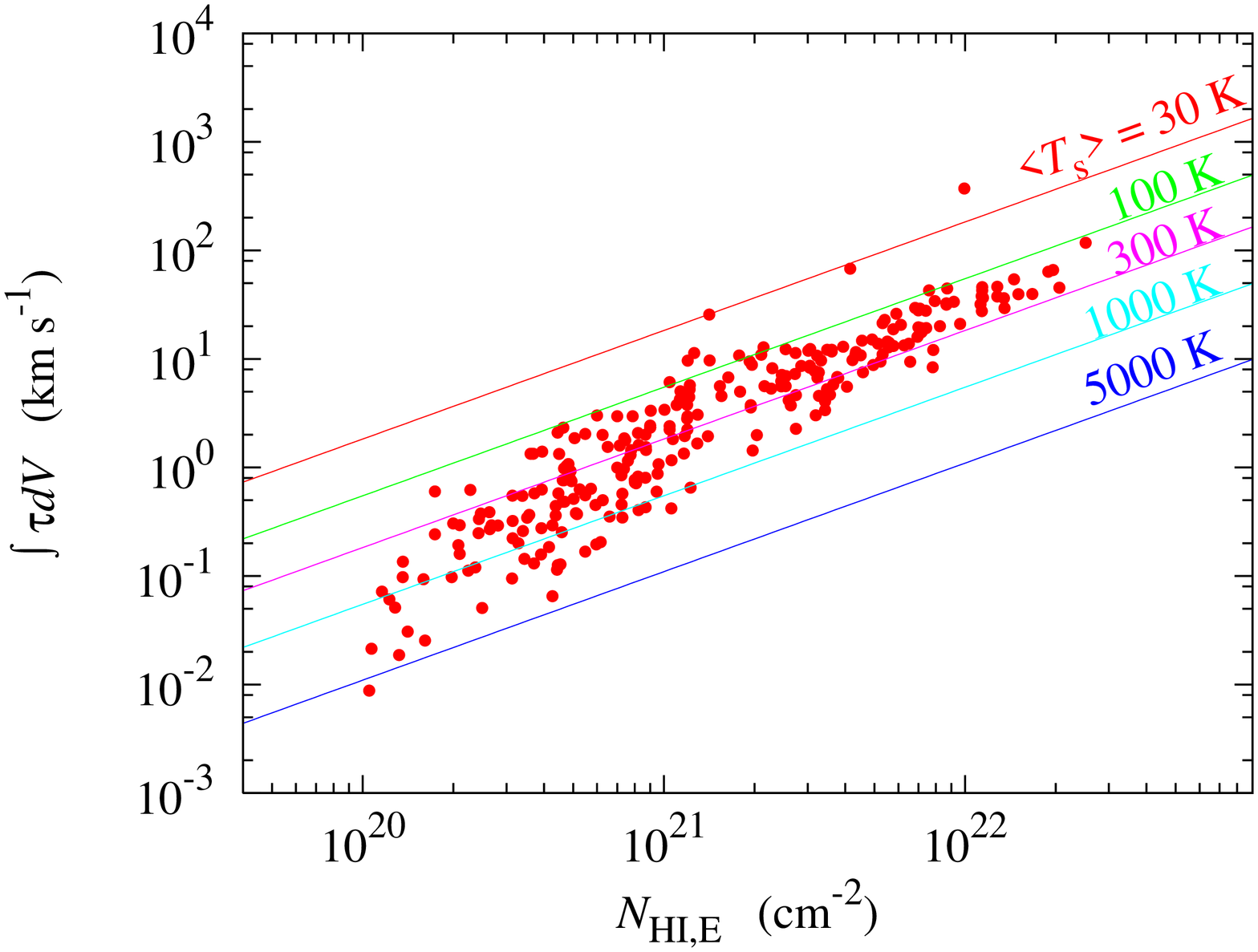}
\caption{Left: A comparison of $\nhies$ and $\nhiot$ for a sample of Galactic 
sightlines. The difference is about $50\%$ for $\nhies \gtrsim 10^{22} \:\:\: 
{\rm cm}^{-2}$. Right: The integrated optical depth as a function of $\nhies$ 
for the same sample. Note that the average $\Ts$ is higher for low $\nhi$ 
sightlines.}
\label{fig:fig7-8} 
\end{center}
\end{figure*}

As shown in the left panel of Figure \ref{fig:fig7-8}, the estimated $\nhi$ is 
$\sim 50\%$ higher than $\nhiot$ when $\nhies \gtrsim 10^{22}\: {\rm cm}^{-2}$. 
Comparing $\nhies$ with $\int \tau \:dV$, the average $\Ts$ for most of these 
sightlines is between $100$ K and $1000$ K, with a trend of a lower $\Ts$ for 
higher $\nhi$, as expected. This is shown in the right panel of Figure 
\ref{fig:fig7-8}. We also see an indication of very low integrated optical 
depth at low $\nhi$ (i.e.~very high $\Ts$ and negligible cold gas fraction), 
suggesting a threshold column density of a few times $10^{20} \: {\rm cm}^{-2}$ 
for cold gas formation. Note that these trends are similar to what have been 
reported earlier by \citet{kanekar11} for a smaller sample. 

Finally, the corrected column density $\nhies$ can be well represented by a 
functional form 
\begin{equation}
\label{eqn:eqnr2}
\nhies = -A\ln\left(1-\frac{\nhiot}{A}\right) \:\:,
\end{equation}
suggested by \citet{st04}. The best fit value of $A = (2.64 \pm 0.06) \times 
10^{22}$ cm$^{-2}$ for our sample is marginally different from the value $2.1 
\times 10^{22}$ cm$^{-2}$ reported for the Galactic plane by \citet{st04}. We 
leave a more detailed analysis of a larger sample to model the observations in 
terms of the temperature of halo and disk gas of the Milky Way for future work.

\section{Discussions and Conclusions}
\label{sec:dscus}

In this work, we have revisited the issue of estimating $\nhi$ from H\,{\sc i} 
$21$ cm absorption and emission spectra. Reliable estimation of $\nhi$ from the 
emission spectra is challenging as the sightlines often pass through mix of 
gases at different temperature. Our knowledge of the relative position of these 
gas clouds is also limited. The issue is even more prominent at higher $\tau$; 
the derived $\nhi$ is significantly biased because the optically thin limit 
underestimates the true $\nhi$. Moreover, suitable continuum background source 
may not be present along the same or nearby sightlines for absorption studies.

We have developed a formalism to get an unbiased estimate of $\nhi$ from only 
the emission spectrum, based on an observed correlation between $\Tb$ and 
$\tau$. The equivalent $\Ts - \tau$ correlation ($\Ts \propto \tau^{-0.43}$) 
from the \citet{roy13a} sample turns out to be in close agreement with that of 
previous studies \citep{lazareff75,heiles03b}. However, to get the $\Ts - \tau$ 
correlation, these studies obtain the peak optical depth $\tau_{0}$ and the 
brightness temperature ${T_{{\rm B,peak}}}$ by modeling the spectrum with 
multiple Gaussians, and the parameters are thus model-dependent. Also, the low 
spectral resolution may lead to ambiguity in determining ${T_{{\rm B,peak}}}$ 
for \citet{lazareff75}. In contrast, our analysis and derived correlation is 
based on directly measured $\Tb(V)$ and $\tau(V)$ from all velocity channels. 
It should be noted that the observed $\Tb-\tau$ correlation only constrain a 
combination of $n$ and $\alpha$, namely $\alpha/(n-\alpha)$. In general, if $n 
\neq 1$, i.e.~the assumption of thermal pressure equilibrium is not valid 
\citep{kulkarniheiles87}, the value of $\alpha$ will depend on $n$. This will, 
however, not affect any of the conclusions as we do not use $\alpha$ or $n$ 
separately in our analysis.

One caveat of the current study is that the $\Tb - \tau$ correlation is derived 
using measurements with very different spatial resolution. A good agreement of 
the observed $\Tb - \tau$ distribution with numerical simulations \citep{kim14} 
indicates the broad consistency of our analysis. However, one would ideally 
like the resolution to be the same (which is practically hard to achieve), or 
systematically study the effect of a larger beam size for the emission spectra 
compared to the absorption spectra. For the complete sample of the absorption 
survey data, we plan to address this in the near future by deriving the $\Tb - 
\tau$ correlation, at least for a sub-sample, using emission spectra at 
different resolution (e.g. LAB survey, Effelsberg and Arecibo telescope data), 
and check how it affects the column density estimation.

The phase fraction distribution also affects the $\nhi$ estimate obtained from 
the emission spectrum. \citet{chengalur13} carried out simulations with many 
different column density and gas temperature distributions to show that 
$\nhiot$ is biased and underestimates $\nhi$, while $\nhits$ is an unbiased 
estimator. Even though their conclusion is qualitatively true, irrespective of 
what $\nhi$ distribution is chosen, the ratio $r=\nhits/\nhiot$ quantitatively 
depends on the phase fraction and column density distributions. Hence, their 
simulation result on the variation of $r$ as a function of $\nhits$ does not 
agree very well with our best fit function from observations, $r = {\rm exp} 
[1.985\times10^{-21}\ {\rm cm^{-2}} \nhits]$, particularly for large $\nhits$ 
values (see Figure \ref{fig:fig2-3}). This is most likely due to the assumption 
that the sightlines pass through a random distribution of gas phases for their 
fiducial case. In reality, the actual phase fraction distribution may be very 
different from a random distribution, and can in principle be derived from the 
observed $\Tb-\tau$ correlation. Finally, once the effect of resolution is 
well-understood, this formalism may be extended for 21 cm observation of other 
galaxies to obtain an unbiased estimate of $\nhi$, as well as to study $\nhi$ 
distribution, power spectra etc. by using only the emission spectrum. 

\section*{Acknowledgements}
We thank the anonymous reviewer for useful comments that helped us improve the 
quality of this manuscript significantly. We also thank S.~Bharadwaj, A.~Sahu 
and J.~N.~Chengalur for their help, and N. Kanekar for valuable suggestions. 
N.~R.~acknowledges support from the Infosys Foundation through the Infosys 
Young Investigator grant.

\bibliographystyle{mn2e}

\bsp

\label{lastpage}

\end{document}